\documentclass[twocolumn]{IEEEtran}
\IEEEoverridecommandlockouts
\usepackage[cmex10]{amsmath}
\usepackage{amssymb, color}
\usepackage{graphicx}
\usepackage{cite}
\usepackage{enumerate}
\graphicspath{{Figures/}}
\usepackage{epsfig}
\newtheorem{defi}{Definition}
\newtheorem{thm}{Theorem}
\newtheorem{lem}{Lemma}
\newtheorem{remark}{Remark}
\newtheorem{example}{Example}

\begin{document}
\title{Diffusion Based Molecular Communication \\ with Limited Molecule Production Rate}
\author{Hamid G. Bafghi, Amin Gohari, Mahtab Mirmohseni, Masoumeh Nasiri-Kenari
\\Department of Electrical Engineering, Sharif University of Technology, Tehran, Iran
}
\maketitle

\begin{abstract}
This paper studies the impact of a transmitter's molecule generation process on the capacity of a concentration based Molecular Communication (MC) system. Constraints caused by the molecule generation process affect the availability of the molecules at the transmitter. The transmitter has a storage of molecules, and should decide whether to release or save the currently produced molecules. As a result, the MC system has conceptual connections with energy harvesting systems. In this paper, we consider two scenarios on the propagation channel. The first scenario assumes a channel with no Inter-symbol Interference (ISI), \emph{i.e.,} a memoryless channel. We derive bounds on the capacity of the MC system in this scenario.
The second scenario assumes the MC network with ISI, in which the output of the channel depends on the history of released molecules in the pervious time-slots.
Based on the assumptions that either the transmitter or the receiver knows the channel statistics, we compute a lower bound on the channel capacity.\footnote{Authors are listed in the alphabetical order, not according to their contributions.}
\end{abstract}

\section{Introduction}
Unlike the classical wireless communication, diffusion based Molecular Communication (MC) utilizes molecules as the carriers of information between the communicating parties. Type, concentration, or the release time of molecules can be used for signaling by a molecular transmitter. As a result, a mechanism must be set in place for production of molecules at the transmitters \cite{bibi168}. For instance, this may be realized by chemical reactions inside the transmitter nodes. The impact of this molecule production process on the capacity of a molecular channel is the subject of this paper.

We assume that the transmitter includes a production unit as well as a storage unit. The production unit adds some amount of new molecules in each time-slot to the storage unit. The production rate of molecules may depend on the amount of molecules already exist in the storage unit, \emph{e.g.,} the chemical process responsible for molecule production might have a faster production rate if the storage unit is empty. The transmitter communicates its message by controlled opening of an outlet of the storage unit for a short period of time and thereby releasing a concentration of molecules into the environment at the beginning of each time-slot. This model is comparable with an \textit{Energy Harvesting} (EH) system in the classical communications, in which the transmitter harvests energy and wishes to send its message such that its transmitted signal is amplitude-constrained to the amount of harvested and stored energy at the transmitter~\cite{bibi146, bibi58, bibi62, bibi64, bibi113, bibi115}.
The transmitter (in the EH system) may have finite~\cite{bibi113} or infinite~\cite{bibi63} energy storage (battery) or it can be assumed with no battery~\cite{bibi138}.

There are several approaches and models of transmitters and receivers for diffusion based MC in the literature. We follow the common approach of choosing one of the models and adapting our analysis to it. In particular, we adopt the macro-scale mode of MC and the molecular Poisson model for our study (see~\cite{bibi162, bibi168} and references therein for a review of different models and the related results). More specifically, we assume that the amount of released molecules is a deterministic concentration (in molar) and the Fick's law of diffusion describes the medium. The reception noise is modeled by a Poisson random variable, \emph{i.e.,} the received signal has a Poisson distribution whose mean is proportional to the average concentration of molecules at the receiver.

In this paper, we consider two cases of the Poisson channel, with or without \textit{Inter-Symbol Interference} (ISI). The ISI occurs when a non-negligible portion of transmitted molecules from the previous time-slots remain in the medium and affect the communication in the current time-slot. We begin by providing a number of capacity results when the channel is without ISI. One should note that though the channel is memoryless in this case, the problem is still complicated due to the fact that the transmitter has memory; the number of released molecules in each transmission is limited by the level of the storage, which itself depends on the previous transmissions. A similar phenomenon occurs in the classical energy harvesting systems. Next, we consider a channel with ISI, \emph{i.e.,} a channel with memory, and provide a result on its capacity depending on whether the channel statistics are known at the transmitter or at the receiver.

\subsection{Related works}
In practice, a MC transmitter suffers from constraints on its molecule production and storage processes~\cite{bibi199, bibi195, bibi200}. The molecule production may be constrained by limitations on the chemical reactions or the availability of food and energy for molecule generation at the transmitter~\cite{bibi200}. Moreover, in practical scenarios, the bio-nanomachines \emph{store} the molecules that they produce internally or capture externally from the environment~\cite{bibi200}. Thus, the limitation on molecule storage forces some constraints on the molecule transmitting process~\cite{bibi202}.

Most of the existing works in the literature of diffusion based MC networks assume availability of a constant number of molecules at the transmission times~\cite{bibi198}. For instance in the on/off keying, the transmitter releases a fixed amount of molecules to send bit `1' and stays silent to send bit `0'~\cite{bibi180, bibi193, bibi197, bibi198, bibi160}. However, there are a few works that consider the limitations on molecule production at the transmitter~\cite{bibi199, bibi195, bibi200}: the random chemical reactions in the molecule production process is considered in~\cite{bibi199}. This work studies the chemical description of the transmitters in terms of Langevin equation. In~\cite{bibi195} the total molecule concentration at the transmitter is assumed to be a function of the number of random chemical reactions between different molecule types. More precisely, the inherent randomness in the availability of food and energy is assumed to affect the molecule generation process which limits the availability of molecules at the transmitter in the beginning of each time-slot~\cite{bibi200}. Due to lack of sufficient molecules, the lengths of the symbol intervals may vary in practical MC systems and the constrained transmitter may not be able to emit molecules with a fixed release frequency ~\cite{bibi201}; the authors in \cite{bibi201} suggest using two types of molecules for communicating and symbol synchronization over the channel.
In~\cite{bibi192}, it is assumed that the amount of available molecules in the present time-slot, which is referred to as the ``state'' of the transmitter, depends on the state and the inputs in the previous time-slots.

In another line of works, the transmitter is assumed to actively capture the needed molecules from the environment.
This process is referred to as ``molecule harvesting''~\cite{bibi202}. Here the transmitter has the ability to harvest the arrived molecules, in addition to the background molecules that randomly hit the transmitter and the ones captured through chemical synthesis from other materials.
In this model, the harvesting process is constrained by the size of the molecule storage at the transmitter. The state of the transmitter (i.e., the number of the molecules at the transmitter's storage) is updated according to a weighted combination of the released, the received and the harvested molecules in the past time-slot~\cite{bibi202}.

In this paper, we study the transmitter molecule production constraint. Our work is novel since previous works do not consider capacity degradation due to the molecule production constraint at the transmitter. For our study, we adopt a molecular Poisson channel model, an important molecular channel adopted in many works in the literature~\cite{bibi203, bibi160, bibi163, bibi161, bibi181, bibi182, bibi186},

The rest of the paper is organized as follows. In Section~\ref{model}, the MC channel and some preliminary definitions are presented.
The main results of the paper on the MC networks with no ISI are presented in~\ref{main_without_ISI}, including the capacity in the case of infinite molecule storage, and inner and outer bounds on the capacity in the case of finite molecule storage.
The main results of the paper on the MC networks with ISI are presented in~\ref{main_with_ISI}. The paper is concluded in Section~\ref{diss}.

\section{System Model}\label{model}
\subsection{Notation and Definitions}
We use the notation $x^n$ to denote the sequence $(x_0, x_1,\ldots, x_n)$. Random variables (r.v.s) are denoted by uppercase letters, while their realizations are denoted by the lowercase letters. We say that random variables $X, Y, Z$ form a Markov chain if $p_{Z|XY}(z|x,y)=p_{Z|Y}(z|y)$ for all $x,y,z$. We show this Markov chain relation by $X\rightarrow Y\rightarrow Z$. 
A sequence of random variables $\{X_k\}$ for $k=1,2,\cdots$ is said to be Asymptotically Mean Stationary (AMS) if 
\begin{equation}
\lim_{n \to \infty} \frac{1}{n} \sum_{k=1}^n \mathbb{P}[X_k \in A]= \overline{P}(A)
\end{equation}
exists for all measurable $A$. Under this condition,  $\overline{P}$ is also a probability measure and is named the \emph{stationary mean} of the AMS sequence \cite{bibi206}.

\subsection{Channel Model}\label{Channel_Model}
In our setting, we have a point-to-point communication channel in which the information is conveyed by the molecule concentration released into the environment by the transmitter. A deterministic channel between the transmitter and the receiver, based on the Fick's law of diffusion, is assumed. A reception noise is assumed at the receiver; the concentration of molecules is detected by a Poisson reception process at the receiver \cite{bibi191}. A common example of the Poisson reception process is the \emph{particle counting noise} of a transparent receiver~\cite{bibi204, bibi205}. A transparent receiver consists of a transparent sphere of a certain volume. It counts the number of molecules that fall into its sphere. A transparent receiver is passive in the sense that it does not affect the diffusion medium by imposing a boundary condition to the differential equation describing the diffusion process. The precise mathematical description of a Poisson reception process is as follows: if the concentration of molecules around the receiver is $\rho$ moles, the receiver's measurement is distributed according to $\textrm{Poisson} (\kappa \rho)$ for some constant $\kappa$. Since the variance of a Poisson distribution is proportional to its mean, the larger the $\rho$, the noisier the receiver's reception will be.

We assume a time-slotted transmission. The transmitter instantaneously releases $X_i$ moles of molecules into the environment at the beginning of each time-slot for $i=0,1,2, \dots$. In other words, if the transmitter is located at $\vec{r}=0$ and has a clock with frequency $1/T_s$, the transmitter's channel input is the impulse train
$$\sum_{k}X_k\delta(\vec{r}=0)\delta(t-kT_s).$$
The Fick's law of diffusion describes propagation of molecules in the environment. Here, we assume that the communication is invariant over time, and the boundary conditions are set to zero, \emph{i.e.,} there is no molecule production source besides the transmitter. As a result, the diffusion medium (described by the Fick's law of diffusion) becomes a linear time-invariant (LTI) system and can be characterized by its impulse response.
Thus, the concentration of molecules at the receiver at the end of the $i$-th time-slot can be expressed as the convolution $\sum_{j=0}^{i} \zeta_{j}X_{i-j}$, where~$\zeta_{j}$ represents the channel impulse response coefficient. The receiver's measurement follows a Poisson distribution as~$Y_i\sim \textrm{Poisson} (\kappa\sum_{j=0}^{i} \zeta_{j}X_{i-j})$. Letting $p_j=\kappa \zeta_j$, we can write $Y_i\sim \textrm{Poisson} (\sum_{j=0}^{i} p_{j}X_{i-j})$ for $i=0,1,2, \dots$.
We refer to $\{p_i\}$ as the channel coefficients.
We assume that the reception noise at the receiver in different time slots are mutually independent, \emph{i.e.,} $Y_i$'s are conditionally independent given the transmission amounts $\{X_i\}$. If the reception noise is  the particle counting noise, to satisfy independence of the reception noises one should adopt the common assumption in the literature that the sampling period $T_s$ is not very small \cite[Section IV]{noel2014optimal}, \cite[Remark 1]{bibi162}.

\begin{figure}
\centering
\epsfig{file=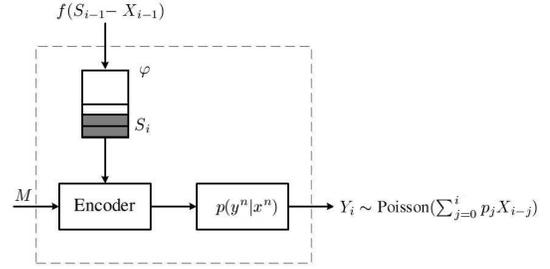,width=.8\linewidth,clip=.5}
\caption{Modeling the molecular communication networks with constrained transmitter and ISI effect as a point-to-point molecule-harvesting Poisson channel with memory.
}
\label{fig:3}
\end{figure}

We say that the channel is without ISI if $p_i$ is negligible for $i>0$. In this case, $Y_i\sim \textrm{Poisson} (p_0 X_i)$, and
\begin{IEEEeqnarray}{rCl}\label{eqn4}
p (y^{n}|x^{n}) =\prod_{i=1}^{n} p(y_{i}|x_{i}).
\end{IEEEeqnarray}
The channel is said to be with ISI if the output of the channel in~$i$-th time-slot depends on the past inputs of the channel in the previous time-slots with weights~$p_{j}, j\in\{0, \ldots, i\} $
(see Fig.~\ref{fig:3}).

\subsection{Transmitter Model}\label{Tx}
The transmitter opens the outlet of its  molecule storage unit for a short period of time at the beginning of each time-slot.
The channel input~$x_i$, at time-slot~$i$, represents the deterministic released molecule concentration (in molar) into the environment.
We assume that the transmitter has~$s_i$ moles of molecules in the beginning of~$i$-th time-slot (in its storage unit).
Moreover, the amount of molecules in the storage unit after recharging for the duration of the~$i$-th time-slot is denoted by $f(s_i)$, where $f(\cdot)$ is a known function.
We expect that $f(s_i)\geq s_i$, \emph{i.e.,} molecule production is non-negative in the~$i$-th time slot. Also, we expect $f(\cdot)$ to be a non-decreasing function, \emph{i.e.,} $f(s)\geq f(s')$ for $s\geq s'$. The intuition here is that if the transmitter starts off with $s$ mole of molecules, it will have more molecules after recharging than if it starts off with $s'<s$ moles of molecules. In some practical scenarios, the production rate of new molecules ($f(s)-s$) decreases in $s$, the amount of molecules already existing in the storage unit. We do not need to make such a restrictive assumption on the function $s\mapsto f(s)-s$ for our results to hold.

If $S_i$ is the amount of molecules in the transmitter at~$i$-th time-slot, we assume that $X_i\leq S_i$ moles of molecules are released into the environment. Thus, we will have
\begin{IEEEeqnarray}{rCl}\label{eqn1}
S_{i+1}=f(S_i-X_i).
\end{IEEEeqnarray}
In other words, $S_i$ reduces to $S_i-X_i$ because of molecule release and then recharges to $f(S_i-X_i)$.

We assume that the transmitter starts off empty, \emph{i.e.,} $S_0=0$. If we have no transmission ($X_i=0$) and just recharging, we will have
$S_i=f(S_{i-1})$ for $i=1,2,\dots$; and thus, $S_i=f^{i}(0)$ where 
$$f^{i}(s)\triangleq (\underbrace{f\circ f\circ f \cdots \circ f}_{i~\text{times}})(s).$$
 Since we assumed that $f(s)\geq s$ for all $s$, the equation $S_i=f(S_{i-1})$ implies that $S_i\geq S_{i-1}$. Therefore, $f^{i}(0)$ is a non-decreasing sequence of non-negative reals. Let
\begin{IEEEeqnarray}{rCl}\label{eqn2}
\varphi=\lim_{i\rightarrow \infty} f^{i}(0),
\end{IEEEeqnarray}
be the amount of molecules after recharging for infinite time (when there is no transmission); we set $\varphi=\infty$ if the sequence $f^{i}(0)$ tends to infinity (as discussed later). From \eqref{eqn2}, it follows that given any $s^*\in[0,\varphi)$, it is possible to save  molecules for a finite number of time instances  $n_0$ and reach the molecular level $s^*$ or larger in the transmitter's reservoir (when there is no transmission). That is $S_{n_0}\geq s^*$.

We say that the transmitter has finite molecule capacity and saturates if $\varphi<\infty$, and has unbounded molecule capacity  if $\varphi=\infty$. If $\varphi<\infty$, from \eqref{eqn2} we have $f(\varphi)=\lim_{i\rightarrow \infty} f^{i+1}(0)=\varphi$, \emph{i.e.,} $\varphi$ is a fixed point of $f$. If $\varphi=\infty$, from monotonicity of $f$ and \eqref{eqn2} we have $$\lim_{x\rightarrow\infty}f(x)=\infty,$$
which we can interpret as $f(\varphi)=\varphi$ for $\varphi=\infty$ by symbolically extending the domain of $f$ to include $\infty$.

With the assumption that the transmitter starts off empty, we only need to define the function $f$ over the interval $[0,\varphi]$. The reason is that for any transmission sequence $\{X_i\}$ satisfying $X_i\leq S_i$, we will have $S_i\leq \varphi$. This follows from induction
\begin{IEEEeqnarray}{rCl}\label{eqn3}
S_{i+1}=f(S_i-X_i)\leq f(S_i)\leq f(\varphi)=\varphi,
\end{IEEEeqnarray}
where we use the monotonicity property of $f(.)$. Hence, we assume that the function $f(.)$ is well-defined over $[0,\varphi]$.

\subsection{Channel Capacity}
The transmitter wishes to send the message~$M$ which is uniformly distributed over the set~$\mathcal{M}= \{0,1\}^{k}$ to the receiver in~$n$ channel uses. The communication rate  is $R=k/n$ bits per channel use. The rate~$R$ is achievable if for any~$0<\epsilon\leq 1$ there exists a code (respecting the transmitter's input constraints) with rate $R-\epsilon$ whose average error probability is less than $\epsilon$. The channel capacity $\mathcal C$ is the maximum achievable communication rate.

Capacity calculation is complicated due to the memory introduced into the problem by the constraints on the transmitter. The transmitter should adapt its molecule release to the amount of produced molecules in the current time-slot and the consumed molecules in the previous time-slots.
Thus, the transmitter has memory.

\section{MC Network Without ISI}\label{main_without_ISI}
In this section, we present our main results on the capacity of the MC network without ISI, \emph{i.e.,}  $Y_i\sim \textrm{Poisson} (p_0 X_i)$ for all $i$.

For a given memoryless channel $p(y|x)$ (here a Poisson channel) on input alphabet $\mathcal{X}=[0,\infty)$ and a real $c\geq 0$, let us define:
\begin{IEEEeqnarray}{rCl}\label{eqn7}
R_c=\sup_{p_X:~\mathbb{E}[X]\leq c} I(X;Y),
\end{IEEEeqnarray}
\begin{IEEEeqnarray}{rCl}\label{eqn8}
\tilde R_c=\sup_{p_X:~0\leq X\leq c} I(X;Y).
\end{IEEEeqnarray}
The difference between $R_c$ and $\tilde{R}_c$ is that in $R_c$ we take supremum over all distributions on $[0,\infty)$ with an expected value be at most $c$, while in $\tilde{R}_c$ we take supremum over all distribution whose support is $[0,c]$.

\begin{thm}\label{thm:a1}
Let
\begin{IEEEeqnarray}{rCl}\label{eqn9}
\Delta_u=\sup_{s\in [0,\varphi]}f(s)-s.
\end{IEEEeqnarray}
Then the channel capacity $\mathcal C$ satisfies:
\begin{IEEEeqnarray}{rCl}\label{eqn10}
\tilde{R}_{\Delta_u}\leq \mathcal C\leq R_{\Delta_u}.
\end{IEEEeqnarray}
\end{thm}
\begin{IEEEproof}
To see the upper bound, note that the number of molecules produced in each recharging period cannot exceed $\Delta_u$. Therefore, the average number of consumed molecules cannot be larger than $\Delta_u$:
\begin{IEEEeqnarray}{rCl}\label{eqn11}
\frac 1n \sum_{i=1}^n X_i\leq \Delta_u.
\end{IEEEeqnarray}
Observe that \eqref{eqn7} is the channel capacity under the average input cost constraint $\Delta_u$ given in \eqref{eqn11}~\cite{bibi69}. As a result, the upper bound then follows from the known result on the capacity of memoryless channels with an average input cost constraint~\cite{bibi69}. Because we assume no ISI in this section, our channel is a memoryless Poisson channel and the result of \cite{bibi69} can be utilized.

It remains to  prove the lower bound on the capacity: $C\geq \tilde{R}_{\Delta_u}$. Take an arbitrary $s\in[0,\varphi)$ and let \begin{align}c=f(s)-s.\label{eqn:a23d}\end{align} We prove that $\tilde R_c=\sup_{p_X:~X\leq c} I(X;Y)$ is an achievable rate. This will complete the proof because ${s}$ has been taken arbitrary in $[0,\varphi)$, and furthermore
$$\Delta_u=\sup_{s\in [0,\varphi]}f(s)-s=\sup_{s\in [0,\varphi)}f(s)-s,$$
as $f(\varphi)-\varphi=0$.
To show that $\tilde R_c$ is achievable, it suffices to show that for any arbitrary distribution $p_X$ on $[0,c]$, the mutual information $I(X;Y)$ is an achievable rate.

Let $s^*=f(s)=s+c$. Then from \eqref{eqn:a23d} we have $f(s^*-c)=s^*$. We claim that $s^*\in[0,\varphi]$. This is because $s^*= f( s)\leq f(\varphi)=\varphi$ where we used the monotonicity property of $f$. Since $s\in[0,\varphi)$, we can use the save strategy at the beginning and wait for a finite number of time instances to reach the molecule level $s$ or larger in the transmitter's reservoir. If we wait for one more time step, we reach  the molecule level $s^*=f(s)$. Hence, there is some finite number $n_0$ such that $S_{n_0}\geq s^*$ assuming that we do not transmit any molecules in the first $n_0$ time steps.

We would like to start transmitting in time instances $i>n_0$. We claim that if we limit the transmission level to $c$, then transmitter's molecular reservoir will not drop below $s^*$. In other words, if $X_i\leq c$ for $i>n_0$ we have $S_i\geq s^*$ for all $i>n_0$. This follows from induction. Assuming that $S_i\geq s^*$ and using the fact that $X_i\leq c$, we have
\begin{IEEEeqnarray}{rCl}\label{eqn12}
S_{i+1}=f(S_i-X_i)\geq f(S_i-c)\geq f(s^*-c)=s^*.
\end{IEEEeqnarray}

The above argument shows that it is possible to transmit \emph{any} sequence $X_i$ satisfying $X_i\leq c$ for all $i>n_0$, as the molecular reservoir is always at least $s^*$ and hence never hits zero. By letting the blocklength $n$ tend to infinity, the initial finite time instances $n_0$ becomes negligible compared to transmission length $n$ and  results in no rate loss. We obtain the achievability of $\tilde R_c=\sup_{p_X:~X\leq c} I(X;Y)$ via Shannon's point to point achievability scheme, \emph{i.e.,} by constructing i.i.d.\ codewords from a $p_X$ defined on $[0,c]$.
\end{IEEEproof}

\begin{thm}\label{thm:a2}
Assume that $\varphi=\infty$. Let
\begin{IEEEeqnarray}{rCl}\label{eqn13}
\Delta_\ell=\liminf_{s\rightarrow\infty}\left(f(s)-s\right).
\end{IEEEeqnarray}
Then the channel capacity can be bounded from below as follows:
\begin{IEEEeqnarray}{rCl}\label{eqn14}
\mathcal C\geq R_{\Delta_\ell}.
\end{IEEEeqnarray}
\end{thm}

\begin{IEEEproof}
We prove that $R_c$ is an achievable rate for any  $c<\Delta_\ell$. By the definition of $\Delta_\ell$, there exists some $s^*>0$ such that for any $s\geq s^*$ we have
\begin{IEEEeqnarray}{rCl}\label{eqn15}
f(s)\geq s+c.
\end{IEEEeqnarray}
To prove that $R_c=\sup_{p_X:~\mathbb{E}[X]\leq c} I(X;Y)$ is achievable, we need to prove achievability of $I(X;Y)$ for an arbitrary distribution $p_X$ satisfying $\mathbb{E}[X]\leq c$. The idea is to mimic  the proof of \cite{bibi64} here and prove this by considering distributions on a finite support. Take some positive real number $K$, and an arbitrary distribution $p_X$ of finite support on $[0,K]$. Here $K$ may be larger than $c$, but the distribution of $X$ is required to satisfy $\mathbb{E}[X]\leq c$. It is shown in \cite{bibi64} that achievability of $I(X;Y)$ for such finite support distributions is enough to show the achievability of $R_c$.

As in~\cite{bibi64}, we wait at the beginning for a finite amount of time and save molecules so that the amount of saved molecules exceeds a number $w$ (to be specified later). Because $\varphi=\infty$, we can take $w$ as large as we want.

As long as $S_i\geq K+s^*$, we have $S_i-X_i\geq s^*$ and hence from \eqref{eqn15} we have
\begin{IEEEeqnarray}{rCl}\label{eqn16}
S_{i+1}=f(S_i-X_i)\geq S_i-X_i+c.
\end{IEEEeqnarray}
In other words, a production of $c$ moles of molecules is guaranteed as long as the molecular reservoir level stays above $K+s^*$.

In \cite{bibi64}, the update rule $S_{i+1}=S_i-X_i+c$ is considered. Here in the inequality \eqref{eqn16}, the sequence $S_i$ has a larger growth and dominates the  process of \cite{bibi64}. This indicates that the same proof can be mimicked here since \eqref{eqn16} shows that in each step, more molecules are produced than the update rule $S_{i+1}=S_i-X_i+c$ of \cite{bibi64}. The number $w$ should be chosen to ensure that with high probability, starting from $w$ molecules, $S_i$ remains above $K+s^*$ so that \eqref{eqn16} remains valid. But this is possible with the same argument as in \cite{bibi64}.
\end{IEEEproof}
\begin{remark}\emph{  Theorems \ref{thm:a1} and \ref{thm:a2} give a tight characterization of capacity $\mathcal C$ for the case of infinite battery capacity $\varphi=\infty$ if $\Delta_\ell=\Delta_u$. For example, this occurs if $f(s)=s+v$ for some constant $v$, \emph{i.e.,} when we have a constant production rate. Theorem \ref{thm:a2} is comparable with an EH system with an infinite battery~\cite{bibi64}. In~\cite{bibi64}, it is shown that the capacity of Gaussian EH channel with an infinite battery, with the average arrived energy~$P$, is equal to the capacity of the Gaussian point-to-point channel with average power constraint~$P$.}
\end{remark}

In the following, we find a lower bound on the capacity when $\varphi<\infty$.

\begin{thm}\label{MC_lower} Assume that $\varphi<\infty$.
Let $q_{X|S}(x|s)$ be a conditional distribution on
$$\{(x,s): f(0)\leq s\leq \varphi, 0\leq x\leq s\}$$
such that the Markov chain
$S_0=0$, $S_{i+1}=f(S_i-X_i)$ with $p_{X_i|S_i}=q_{X|S}$ asymptotically converges to a stationary distribution
$q_S$, \emph{i.e.,} $p_{S_i}$ tends to $q_S$  in total variation distance as $i$ tends to infinity.  Then the capacity is lower bounded as
\begin{IEEEeqnarray}{rCl}\label{eqn19}
\mathcal{C}\geq I(X;Y|S),
\end{IEEEeqnarray}
where $q_{X,S,Y}=q_{X,S}p_{Y|X}$ where $p_{Y|X}$ is the Poisson channel $Y\sim \textrm{Poisson} (p_0 X)$.
\end{thm}
\begin{IEEEproof}
Assume that we start with $S_0=0$ and choose $X_i$ from $S_i$ according to $q_{X|S}$. Consider the update rule
$S_{i+1}=f(S_i-X_i).$
This yields  a sequence $\{(S_i, X_i,)\}$ for $i=0,1,2,\cdots$. Let $Y_i\sim \textrm{Poisson} (p_0 X_i)$ be a memoryless channel. Then we have
\begin{IEEEeqnarray}{rCl}\label{eqnnew21}
p(y^n,x^n, s^n)=\prod_{i=1}^n p(s_i|x_{i-1},s_{i-1})q(x_i|s_i)p(y_i|x_i).
\end{IEEEeqnarray}
We claim that $\{(S_i, X_i, Y_i)\}$ is an Asymptotically Mean Stationary (AMS) sequence. As argued in \cite{bibi206}, the AMS condition allows us to conclude that
$$\lim_{n\rightarrow\infty} \frac{1}{n}I(X^n;Y^n)$$
is a lower bound on the capacity of the channel. Next, we show that $\{(S_i, X_i, Y_i)\}$ is an Asymptotically Mean Stationary (AMS) sequence. Observe that $\{S_i\}$ is a Markov chain (on a countable state space) starting from $S_0=0$. Furthermore, from the definition of the $q_{S,X}$ in the statement of the theorem, $q_S$ is a limiting stationary distribution for this Markov chain and $p(S_i)$ converges to $q_S$ in total variation distance. Because $p_{X_i,Y_i|S_i}(x_i,y_i|s_i)=q_{X|S}(x_i|s_i)p_{Y|X}(y_i|x_i)$, the distribution of $p_{S_i, X_i, Y_i}$ also converges to $q_{S,X,Y}$ and $\{(S_i, X_i, Y_i)\}$ will be proven to be an Asymptotically Mean Stationary (AMS) sequence.

To sum this up, we have that
$$\lim_{n\rightarrow\infty} \frac{1}{n}I(X^n;Y^n)$$
is a lower bound on the capacity of the channel.
Next, note that
\begin{IEEEeqnarray}{rCl}\label{eqn21}
I(X^n;Y^n)&=&H(Y^n)-H(Y^n|X^n)\IEEEnonumber\\
&=& H(Y^n)-\sum_{i=1}^nH(Y_i|X_i)\IEEEnonumber\\
&=& \sum_{i=1}^n H(Y_i|Y_{1:i-1})-H(Y_i|X_i)\IEEEnonumber\\
&\geq& \sum_{i=1}^n H(Y_i|Y_{1:i-1},S_i)-H(Y_i|X_i)\IEEEnonumber\\
&=&\sum_{i=1}^n H(Y_i|S_i)-H(Y_i|X_i)\IEEEnonumber\\
&=&\sum_{i=1}^n H(Y_i|S_i)-H(Y_i|X_i,S_i)\IEEEnonumber\\
&=&\sum_{i=1}^nI(X_i;Y_i|S_i)\IEEEnonumber
\end{IEEEeqnarray}
where we used the fact that $Y_i\rightarrow X_i\rightarrow S_i$ and
$Y_i\rightarrow S_i\rightarrow Y_{1:i-1}$ form Markov chains, which are
 implied from \eqref{eqnnew21}. Because $p_{S_i, X_i, Y_i}$ tends to $q_{S,X,Y}$ as $i$ tends to infinity, we have
$$\lim_{n\rightarrow\infty} \frac{1}{n}I(X^n;Y^n)\geq \lim_{n\rightarrow\infty} \frac{1}{n}\sum_{i=1}^nI(X_i;Y_i|S_i)=I_q(X;Y|S).$$

\end{IEEEproof}


\section{MC Network With ISI}\label{main_with_ISI}
In this section, we present our main results on the capacity of the MC network with ISI. As we mentioned in Section~\ref{Channel_Model}, in this channel~$X_i$ and~$Y_i\sim \textrm{Poisson} (\sum_{j=0}^{i} p_{j}X_{i-j})$ represent the channel input and the channel output, respectively (see Fig.~\ref{fig:3}). Let $\mathcal{C}(p)$ denote the capacity of  a channel with coefficient sequence $p$. Calculation of $\mathcal{C}(p)$ is complicated by the fact that the channel has memory.

Assume that the channel coefficient sequence $p$ belong to a class $\mathcal{P}$. Here, we consider two scenarios: the first scenario assumes that the  transmitter is aware of the actual $p\in\mathcal{P}$, but the receiver only knows the class $\mathcal{P}$ and is unaware of the sequence $p\in\mathcal{P}$. A rate is achievable if there is a sequence of coding strategies for the transmitter and receiver whose error probability vanishes regardless of the choice of $p\in\mathcal{P}$. The second scenario is the other way around; it assumes that the  receiver is aware of the actual $p\in\mathcal{P}$, and the transmitter only knows the class $\mathcal{P}$. Thus, the two scenarios differ in terms of whether the ``channel state information" is available at the transmitter or receiver.

To state our results, we begin by a definition:

\begin{defi}Given two sequences $p=(p_j)$ and $\tilde p=(\tilde p_j)$ of channel coefficients, we say $\tilde p\preceq  p$  if
\begin{IEEEeqnarray}{rCl}
 \tilde p_j= \sum_{k=0}^j q_{j-k}  p_k,
\end{IEEEeqnarray}\color{black}
for some non-negative sequence $q_0, q_1, \cdots$ satisfying
$\sum_{j=0}^\infty q_{i}\leq 1$.
\end{defi}
Intuitively speaking, $\tilde p\preceq  p$ can be understood as a channel with a coefficient sequence $\tilde p$ having a more spread out channel coefficient profile than the channel with a coefficient sequence $p$.
\begin{example}\label{example1}
As an example, given any $\tilde p$ we have $\tilde p\preceq  p$ where $p=(p_j)$ is defined as follows: $p_i=0$ for $i\geq 1$ and $p_0$ is any number greater than or equal to $\sum_{i}\tilde p_i$.
\end{example}

Our main results are as follows:

\begin{thm}\label{thm4} Assume that the molecule production function $f(\cdot)$ is a concave function (in addition to the constraints of Section \ref{Tx}).
Take some channel coefficient sequence $\tilde p=(\tilde p_j)$ such that  $\tilde p\preceq p$ for all $p\in\mathcal{P}$. Then, $\mathcal{C}(\tilde{p})$ is an achievable rate for any channel coefficient profile $p\in\mathcal{P}$ when $p$ is known \underline{only at the transmitter}.
\end{thm}

\begin{thm}\label{thm5}
Take some channel coefficient sequence $\tilde p=(\tilde p_j)$ such that  $\tilde p\preceq p$ for all $p\in\mathcal{P}$. Then, for any molecule production function $f(\cdot)$, $\mathcal{C}(\tilde{p})$ is an achievable rate  for any channel coefficient profile $p\in\mathcal{P}$ when $p$ is known \underline{only at the receiver}.
\end{thm}

\begin{remark} Theorem \ref{thm4} imposes more restrictions on the function $f(\cdot)$ compared to Theorem \ref{thm5}. Functions $f(x)=\min(x+c,\varphi)$ and $f(x)=\sqrt{x+c}$ for $c>0$ are two examples of increasing and concave functions.
\end{remark}
\begin{remark}\label{remark3}  From Example \ref{example1}, we know that given any channel coefficient sequence $\tilde p=(\tilde p_j)$, the sequence $p=(\sum_{j}\tilde p_j,0,0,\cdots, 0)$ satisfies $\tilde p\preceq p$. Since ${p}$ corresponds to a memoryless channel, the above theorems confirms that ISI cannot be utilized to increase the channel capacity.

\end{remark}
\begin{IEEEproof}[Proof of Theorem \ref{thm4}]
In this case, the receiver is unaware of the $p\in\mathcal{P}$. The receiver proceeds with the assumption that the true channel coefficient is $\tilde{p}$. The idea is to show that any code for a channel with coefficient $\tilde{p}$ can be ``simulated" by a code for channel $p$ at the transmitter. More specifically, choose an arbitrary $n$-code of rate $R$ designed for a channel with coefficients $\tilde{p}$ consisting of a set of codewords~$\tilde x^n(m)$ for $m\in\{1, 2, \ldots, 2^{nR}\}$. The codeword $\tilde x^n(m)$ satisfies the transmitter molecule production constraint (i.e., transmission in each stage is less than the amount of molecules available in the transmitter reservoir).
The receiver assumes that the actual channel coefficient is $\tilde{p}$ and uses the decoding algorithm for this channel code. The transmitter gets the true channel coefficient $p$ and uses the following strategy:
\begin{enumerate}
\item From $\tilde p\preceq p$, it computes a non-negative sequence  $(q_0, q_1, \ldots)$ such that
$$ \tilde p_{j}= \sum_{k=0}^j q_{j-k} p_{k}.$$
\item To transmit message $m\in\{1, 2, \ldots, 2^{nR}\}$, it transmits
\begin{align}{x}_{j}(m)= \sum_{k=0}^j q_{j-k} \tilde x_{k}(m), j=0,1, \ldots, n,\label{dsd}\end{align}
on the channel, where $\tilde x^n(m)$ is the codeword of the codebook for $\tilde{p}$ corresponding to message $m$.
\end{enumerate}
With this strategy, the receiver gets $Y_i\sim \textrm{Poisson} (\sum_{j=0}^{i}  p_{i-j} x_{j}(m))$. Observe that
\begin{IEEEeqnarray}{rCl}\label{eqn61}
\sum_{j=0}^{i} p_{i-j} x_{j}(m)
&=& \sum_{j=0}^{i} p_{i-j}\sum_{k=0}^j q_{j-k} \tilde x_{k}(m)
\\&=& \sum_{k=0}^{i}  \tilde x_{k}(m) \Big(\sum_{j=k}^i q_{j-k}p_{i-j} \Big)\IEEEnonumber
\\&=& \sum_{k=0}^{i}  \tilde x_{k}(m) \tilde p_{i-k}.\IEEEnonumber
\end{IEEEeqnarray}
Therefore, $Y_i\sim \textrm{Poisson} (\sum_{j=0}^{i}  \tilde p_{i-j} \tilde x_{j}(m))$, as if the codeword $\tilde x^n(m)$ was transmitted over a channel with the coefficient sequence  $\tilde{p}$. Hence, \eqref{dsd} allows for the simulation of channel $\tilde{p}$ from channel $p$.

The crucial and more difficult step is to show that the sequence of transmission $x^n(m)$ defined in \eqref{dsd}  satisfies the transmitter cost constraint. To prove this, we use the assumption that each codeword $\tilde x^n(m)$ satisfies the transmitter cost constraint. That is, $$\tilde x_i(m)\leq \tilde s_i(m)$$
where $\tilde s_0(m)=0$ and $\tilde s_i(m)=f(\tilde s_{i-1}(m)-\tilde x_{i-1}(m))$ is the number of molecules in transmitter's reservoir at time $i$ if message $m$ is transmitted. Our claim follows  from Lemma \ref{Lemma:AMn} in the Appendix.
\end{IEEEproof}

\begin{IEEEproof}[Proof of Theorem \ref{thm5}]
In this case, the transmitter is unaware of the $p\in\mathcal{P}$. The transmitter proceeds with the assumption that the true channel coefficient sequence is $\tilde{p}$. The idea is to convert the channel
$p$ to the channel $\tilde{p}$ by an operation at the receiver.
Choosing an arbitrary $n$-code of rate $R$ designed for a channel with a coefficient sequence $\tilde{p}$ consisting of a set of codewords~$\tilde x^n(m)$ for $m\in\{1, 2, \ldots, 2^{nR}\}$, the transmitter sends the codeword $\tilde x^n(m)$, assuming that the actual channel coefficient sequence is $\tilde{p}$. The receiver gets the true channel coefficient sequence $p$ and uses the following strategy: from $\tilde p\preceq p$, the receiver computes a non-negative sequence  $(q_0, q_1, \ldots)$ such that
$$ \tilde p_{j}= \sum_{k=0}^j q_{k} p_{j-k}.$$
Next, having received the sequence $(Y_0, Y_1, Y_2, \cdots, Y_n)$, the receiver produces the sequence   $(\tilde Y_0, \tilde Y_1, \tilde Y_2, \cdots, \tilde Y_n)$ as follows: for each $0\leq k\leq n$, we produce random variables $\hat{Y}_{0k}, \hat{Y}_{1k}, \hat{Y}_{2k}, \cdots, \hat{Y}_{kk}, Z_k$ from a multinomial distribution with $Y_k$ balls and probability sequence $(q_k, q_{k-1}, \cdots, q_0, 1-\sum_{i=0}^kq_i)$. We assume that the variables $\hat{Y}_{0k}, \hat{Y}_{1k}, \hat{Y}_{2k}, \cdots, \hat{Y}_{kk}, Z_k$ for different  values of $k$ are produced independently. Then, we let
$$\tilde{Y}_j=\sum_{i=0}^j \hat{Y}_{ij}.$$
Note that $Y_i\sim\textrm{Poisson} (\sum_{k=0}^{i} p_{i-k}\tilde x_{k}(m)),0\leq i\leq n$ are independent  given $\tilde x^n(m)$. By the thinning property of Poisson distribution, and the fact that sum of independent Poisson random variables is again a Poisson random variable, we obtain that $\hat{Y}_{ij}\sim\textrm{Poisson} (q_{j-i}\sum_{k=0}^{i} p_{i-k}\tilde x_{k}(m))$ and furthermore, $\hat{Y}_{ij}$ for different values of $i,j$ are mutually independent. We have
\begin{align*}\tilde{Y}_j=\sum_{i=0}^j \hat{Y}_{ij}&\sim\textrm{Poisson} (\sum_{i=0}^jq_{j-i}\sum_{k=0}^{i} p_{i-k}\tilde x_{k}(m))
\\&=\textrm{Poisson} (\sum_{k=0}^{j}\tilde x_{k}(m)\sum_{i=k}^jq_{j-i} p_{i-k})
\\&=\textrm{Poisson} (\sum_{k=0}^{j}\tilde x_{k}(m)\tilde{p}_{j-k}).
\end{align*}
Furthermore, $\tilde{Y}_j$ for different values of $j$ are mutually independent because they involve summation over disjoint sets of $\hat{Y}_{ij}$. As a result, $\tilde{Y}_i$ is exactly what we obtain if $\tilde x^n(m)$ were passed through the channel $\tilde{p}$. This completes the proof of the receiver being able to apply a post-processing to convert the channel $p$ to $\tilde{p}$.

\end{IEEEproof}

\section{Conclusions}\label{diss}
In this paper, a molecular communication (MC) system is considered in which the information is encoded in the concentration of the molecules, and the molecule generation process causes some constraints on the transmitter.
Moreover, it is assumed that the number of received molecules at the receiver follows a Poisson distribution of the channel input.

Two scenarios on the MC channel with no Inter-symbol Interference (ISI) and the MC channel with ISI were studied. For the case of no-ISI scenario, lower and upper bounds on the channel capacity were derived. For the case of channels with ISI, we provided a capacity result for the cases where the channel coefficient sequence is known either at the transmitter or at the receiver.

\bibliography{thesisbib}

\begin{thebibliography}{10}
\providecommand{\url}[1]{#1}
\csname url@samestyle\endcsname
\providecommand{\newblock}{\relax}
\providecommand{\bibinfo}[2]{#2}
\providecommand{\BIBentrySTDinterwordspacing}{\spaceskip=0pt\relax}
\providecommand{\BIBentryALTinterwordstretchfactor}{4}
\providecommand{\BIBentryALTinterwordspacing}{\spaceskip=\fontdimen2\font plus
\BIBentryALTinterwordstretchfactor\fontdimen3\font minus
  \fontdimen4\font\relax}
\providecommand{\BIBforeignlanguage}[2]{{%
\expandafter\ifx\csname l@#1\endcsname\relax
\typeout{** WARNING: IEEEtran.bst: No hyphenation pattern has been}%
\typeout{** loaded for the language `#1'. Using the pattern for}%
\typeout{** the default language instead.}%
\else
\language=\csname l@#1\endcsname
\fi
#2}}
\providecommand{\BIBdecl}{\relax}
\BIBdecl

\bibitem{bibi168}
N.~Farsad, H.~B. Yilmaz, A.~Eckford, C.~B. Chae, and W.~Guo, ``A comprehensive
  survey of recent advancements in molecular communication,'' \emph{IEEE
  Commun. Surv. Tutorials}, vol.~18, no.~3, pp. 1887--1919, thirdquarter 2016.

\bibitem{bibi146}
M.~L. Ku, W.~Li, Y.~Chen, and K.~J.~R. Liu, ``Advances in energy harvesting
  communications: Past, present, and future challenges,'' \emph{IEEE Comm.
  Surveys Tutorials}, vol.~18, no.~2, pp. 1384--1412, Secondquarter 2016.

\bibitem{bibi58}
O.~Ozel, J.~Yang, and S.~Ulukus, ``Broadcasting with a battery limited energy
  harvesting rechargeable transmitter,'' in \emph{Int. Symp. on Modeling and
  Optimization in Mobile, Ad Hoc and Wireless Networks (WiOpt)}, 2011, pp.
  205--212.

\bibitem{bibi62}
R.~Rajesh, V.~Sharma, and P.~Viswanath, ``Information capacity of energy
  harvesting sensor nodes,'' in \emph{Int. Symp. on Inf. Theory (ISIT)}, July
  2011, pp. 2363--2367.

\bibitem{bibi64}
O.~Ozel and S.~Ulukus, ``Achieving \textsc{AWGN} capacity under stochastic
  energy harvesting,'' \emph{IEEE Trans. on Inf. Theory}, vol.~58, no.~10, pp.
  6471--6483, Oct. 2012.

\bibitem{bibi113}
K.~Tutuncuoglu and A.~Yener, ``Optimum transmission policies for battery
  limited energy harvesting systems,'' \emph{IEEE Trans. Wireless Comm.},
  vol.~11, no.~3, pp. 1180--1189, Mar. 2012.

\bibitem{bibi115}
S.~Ulukus, A.~Yener, E.~Erkip, O.~Simeone, M.~Zorzi, P.~Grover, and K.~Huang,
  ``Energy harvesting wireless communications: a review of recent advances,''
  \emph{IEEE Jour. Selec. Area Comm.}, vol.~33, no.~3, pp. 360--381, Mar. 2015.

\bibitem{bibi63}
O.~Ozel and S.~Ulukus, ``Information-theoretic analysis of an energy harvesting
  communication system,'' in \emph{Int. Symp. on Personal, Indoor and Mob.
  Radio Comm. Workshops (PIMRC Workshops)}, Sept. 2010, pp. 330--335.

\bibitem{bibi138}
H.~G. Bafghi, B.~Seyfe, M.~Mirmohseni, and M.~R. Aref, ``Capacity of channel
  with energy harvesting transmitter,'' \emph{IET Comm.}, vol.~9, no.~4, pp.
  526--531, 2015.

\bibitem{bibi162}
A.~Gohari, M.~Mirmohseni, and M.~Nasiri-Kenari, ``Information theory of
  molecular communication: Directions and challenges,'' \emph{IEEE Trans. on
  Molecular, Biological and Multi-Scale Commun.}, vol.~2, no.~2, pp. 120--142,
  Dec 2016.

\bibitem{bibi199}
D.~T. Gillespie, ``The chemical langevin equation,'' \emph{The Journal of
  Chemical Physics}, vol. 113, 2000.

\bibitem{bibi195}
M.~Pierobon and I.~F. Akyildiz, ``A statistical--physical model of interference
  in diffusion-based molecular nanonetworks,'' \emph{IEEE Trans. on Commun.},
  vol.~62, 6 2014.

\bibitem{bibi200}
B.~Alberts, D.~Bray, K.~Hopkin, A.~Johnson, J.~Lewis, M.~Raff, K.~Roberts, and
  P.~Walter, \emph{Essential Cell Biology}, 4th~ed.\hskip 1em plus 0.5em minus
  0.4em\relax Garland Science, 2014.

\bibitem{bibi202}
J.-T. Huang and C.-H. Lee, ``On capacity bounds of two-way diffusion channel
  with molecule harvesting,'' in \emph{IEEE Int. Conf. on Commun. (ICC), Paris,
  France, 2017}, 2017.

\bibitem{bibi198}
M.~Pierobon and I.~F. Akyildiz, \emph{Fundamentals of Diffusion--Based
  Molecular Communication in Nanonetworks}.\hskip 1em plus 0.5em minus
  0.4em\relax Now Publisher, 2014.

\bibitem{bibi180}
R.~Mosayebi, H.~Arjmandi, A.~Gohari, M.~Nasiri-Kenari, and U.~Mitra,
  ``Receivers for diffusion-based molecular communication: Exploiting memory
  and sampling rate,'' \emph{IEEE Journal on Selected Areas in Commun.},
  vol.~32, no.~12, pp. 2368--2380, Dec 2014.

\bibitem{bibi193}
C.~Jiang, Y.~Chen, and K.~J. Ray~Liu, ``Nanoscale molecular communication
  networks: a game-theoretic perspective,'' \emph{EURASIP Jour. on Adv. in
  Signal Proc.}, vol.~5, 2015.

\bibitem{bibi197}
J.~Suzuki, T.~Nakano, and M.~J. Moore, \emph{Modeling, Methodologies and Tools
  for Molecular and Nano-scale Communications}.\hskip 1em plus 0.5em minus
  0.4em\relax Springer Int. Publishing, 2017.

\bibitem{bibi160}
H.~Arjmandi, M.~Movahednasab, A.~Gohari, M.~Mirmohseni, M.~Nasiri-Kenari, and
  F.~Fekri, ``Isi-avoiding modulation for diffusion-based molecular
  communication,'' \emph{IEEE Trans. on Molecular, Biological and Multi-Scale
  Commun.}, vol.~3, no.~1, pp. 48--59, March 2017.

\bibitem{bibi201}
V.~Jamali, A.~Ahmadzadeh, and R.~Schober, ``Symbol synchronization for
  diffusive molecular communication systems,'' in \emph{IEEE Int. Conf. on
  Commun. (ICC), Paris, France}, 2017.

\bibitem{bibi192}
T.~Nakano, T.~Suda, Y.~Okaie, M.~J. Moore, and A.~V. Vasilakos, ``Molecular
  communication among biological nanomachines: A layered architecture and
  research issues,'' \emph{IEEE Trans. on NanoBioscience}, vol.~13, 9 2014.

\bibitem{bibi203}
M.~Pierobon and I.~F. Akyildiz, ``Capacity of a diffusion-based molecular
  communication system with channel memory and molecular noise,'' \emph{IEEE
  Trans. on Inf. Theory}, vol.~59, 02 2013.

\bibitem{bibi163}
G.~Aminian, H.~Arjmandi, A.~Gohari, M.~Nasiri-Kenari, and U.~Mitra, ``Capacity
  of diffusion-based molecular communication networks over lti-poisson
  channels,'' \emph{IEEE Trans. on Molecular, Biological and Multi-Scale
  Communications}, vol.~1, no.~2, pp. 188--201, June 2015.

\bibitem{bibi161}
S.~M. Mustam, S.~K. Syed-Yusof, and S.~Zubair, ``Capacity and delay spread in
  multilayer diffusion-based molecular communication (dbmc) channel,''
  \emph{IEEE Trans. on NanoBioscience}, vol.~15, no.~7, pp. 599--612, Oct 2016.

\bibitem{bibi181}
J.~Sun and H.~Li, ``On the capacity of amplitude modulation based molecular
  communication channels,'' in \emph{2016 8th Int. Conf. on Wireless Commun.
  Signal Proc. (WCSP)}, Oct 2016, pp. 1--6.

\bibitem{bibi182}
B.~Atakan and O.~B. Akan, \emph{On Channel Capacity and Error Compensation in
  Molecular Communication}.\hskip 1em plus 0.5em minus 0.4em\relax Berlin,
  Heidelberg: Springer Berlin Heidelberg, 2008, pp. 59--80.

\bibitem{bibi186}
G.~Aminian, M.~Farahnak-Ghazani, M.~Mirmohseni, M.~Nasiri-Kenari, and F.~Fekri,
  ``On the capacity of point-to-point and multiple-access molecular
  communications with ligand-receptors,'' \emph{IEEE Trans. on Molecular,
  Biological and Multi-Scale Commun.}, vol.~1, no.~4, pp. 331--346, Dec 2015.

\bibitem{bibi191}
H.~Mahdavifar and A.~Beirami, ``Diffusion channel with poisson reception
  process: capacity results and applications,'' in \emph{2015 IEEE Int. Symp.
  on Inf. Theory (ISIT)}, June 2015, pp. 1956--1960.

\bibitem{bibi204}
D.~Kilinc and O.~B. Akan, ``Receiver design for molecular communication,''
  \emph{IEEE Journal on Selected Areas in Communications}, vol.~31, no.~12, pp.
  705--714, 2013.

\bibitem{bibi205}
M.~Pierobon and I.~F. Akyildiz, ``Diffusion-based noise analysis for molecular
  communication in nanonetworks,'' \emph{IEEE Transactions on Signal
  Processing}, vol.~59, no.~6, pp. 2532--2547, 2011.

\bibitem{noel2014optimal}
A.~Noel, K.~C. Cheung, and R.~Schober, ``Optimal receiver design for diffusive
  molecular communication with flow and additive noise,'' \emph{IEEE
  transactions on nanobioscience}, vol.~13, no.~3, pp. 350--362, 2014.

\bibitem{bibi69}
A.~El~Gamal and Y.-H. Kim, \emph{Network Information Theory}.\hskip 1em plus
  0.5em minus 0.4em\relax Cambridge University Press, 2011.

\bibitem{bibi206}
R.~Rajesh, V.~Sharma, and P.~Viswanath, ``Capacity of {Gaussian} channels with
  energy harvesting and processing cost,'' \emph{IEEE Transactions on
  Information Theory}, vol.~60, no.~5, pp. 2563--2575, May 2014.

\end{thebibliography}
\bibliographystyle{IEEEtran}

\appendix

\begin{lem}\label{Lemma:AMn}
Let $(\tilde x_0, \tilde s_0, \tilde x_1, \tilde s_1, \cdots)$ be a sequence satisfying $\tilde s_0=0$, $\tilde x_i\leq \tilde s_i$,
and $\tilde s_{i+1}=f(\tilde s_{i}-\tilde x_{i})$ for $i\geq 0$. Let $q_0, q_1, \cdots$ be a sequence of non-negative numbers satisfying $\sum_{i}q_i\leq 1$. Let
$${x}_{j}= \sum_{k=0}^j q_{j-k} \tilde x_{k}, \qquad j=0,1,\ldots.$$
Then, assuming that $f$ is a concave molecule production function, there is a sequence $(s_0,  s_1, \cdots)$ of non-negative numbers satisfying
 $s_0=0$, ${x}_i\leq {s}_i$ and $s_{i+1}=f(s_{i}-x_{i})$ for any $i\geq 0$.
\end{lem}

\begin{IEEEproof}[Proof of Lemma \ref{Lemma:AMn}]
We begin by first proving the following fact about the function $f$: given any non-negative numbers $\alpha_i$ and $v_i$ such that $\sum_i\alpha_i\leq 1$ we have
\begin{align}f(\sum_i\alpha_iv_i)\geq \sum_i\alpha_if(v_i).\label{defff}\end{align}
To see this, let $\kappa=\sum_j \alpha_j\in[0,1]$.  Then,
\begin{align}f(\sum_i\alpha_iv_i)&=f((1-\kappa)\cdot 0+\sum_i\alpha_iv_i)\nonumber
\\&\geq (1-\kappa)f(0)+\sum_i \alpha_i f(v_i)\label{eqasa1}
\\&\geq \sum_i\alpha_if(v_i),\nonumber
\end{align}
where \eqref{eqasa1}  from Jensen's inequality for the concave function $f$.

We prove the Lemma by induction on $i$. We form the induction by assuming a stronger induction hypothesis:
we assume that we have defined ${s}_i$ for $i\leq T$ such that ${x}_i\leq {s}_i$, $ s_i=f(s_{i-1}- x_{i-1})$ and additionally
\begin{align}{s}_{j}\geq \sum_{k=0}^j q_{j-k} \tilde s_{k}, \qquad j=1,2, \ldots, T.\label{eqfdfdf343}\end{align}
Then, we prove that ${s}_{T+1}=f(s_{T}- x_{T})$ will satisfy
\begin{align}{x}_{T+1}&\leq {s}_{T+1}\label{eqnd1e1}\\
{s}_{T+1}&\geq \sum_{k=0}^{T+1} q_{T+1-k} \tilde s_{k}.\label{eqnd1e2}\end{align}
The base of the induction is clear.
Since $\tilde x_1\leq \tilde s_1=0$, we have ${x}_1=q_0\tilde x_1=0$. Therefore, ${x}_1\leq {s}_1=0$ and ${s}_{1}\geq q_0 \tilde s_{1}$ hold.
Equation \eqref{eqnd1e2} holds because
\begin{align}{s}_{T+1}&=f(s_{T}-x_{T})\nonumber
\\&\geq f(\sum_{k=0}^{T} q_{T-k} \tilde s_{k}- x_{T})\label{eqasa3}
\\&= f(\sum_{k=0}^{T} q_{T-k} (\tilde s_{k}-\tilde x_{k}))\label{eqasa34}
\\&\geq \sum_{k=0}^{T} q_{T-k}f(\tilde s_{k}-\tilde x_{k})\label{eqasa4}
\\&= \sum_{k=0}^{T} q_{T-k}\tilde s_{k+1}\label{eqasa5}
\\&= q_{T+1}\tilde s_0+\sum_{k=0}^{T} q_{T-k}\tilde s_{k+1}\label{eqasa6}
\\&=\sum_{k=0}^{T+1} q_{T+1-k} \tilde s_{k},\nonumber
\end{align}
where \eqref{eqasa3} follows from the induction hypothesis and monotonicity of $f$, \eqref{eqasa34} follows from the definition of ${x}_T$, \eqref{eqasa4} follows from \eqref{defff}, \eqref{eqasa5} follows from the definition of $\tilde s_{k+1}$ and \eqref{eqasa6} follows from the fact that $\tilde s_0=0$.

Equation \eqref{eqnd1e1} holds because
\begin{align}{x}_{T+1}&=
\sum_{k=0}^{T+1} q_{T+1-k}\tilde x_{k}\nonumber
\\&=\sum_{k=1}^{T+1} q_{T+1-k}\tilde x_{k}\label{eqasadff6}
\\&\leq \sum_{k=1}^{T+1} q_{T+1-k}\tilde s_{k}\label{eqasadff7}
\\&= \sum_{k=1}^{T+1} q_{T+1-k}f(\tilde s_{k-1}-\tilde x_{k-1})\label{eqasadff8}
\\&\leq f(\sum_{k=1}^{T+1} q_{T+1-k}(\tilde s_{k-1}-\tilde x_{k-1}))\label{eqasadff9}
\\&= f(\sum_{k=0}^{T} q_{T-k}(\tilde s_{k}-\tilde x_{k}))\nonumber
\\&= f\bigg(\big(\sum_{k=0}^{T} q_{T-k}\tilde s_{k}\big)-{x}_{T}\bigg)\label{eqasadff11}
\\&\leq f({s}_T-{x}_{T})\label{eqasadff12}
\\&= {s}_{T+1},\nonumber
\end{align}
where \eqref{eqasadff6} follows from $\tilde x_0=0$, \eqref{eqasadff7} follows from $\tilde x_k\leq \tilde s_k$, \eqref{eqasadff8} follows from the definition of $\tilde s_k$, \eqref{eqasadff9} follows from \eqref{defff}, \eqref{eqasadff11} follows from definition of ${x}_T$ and finally, \eqref{eqasadff12} follows from induction hypothesis \eqref{eqfdfdf343} and the fact that $f$ is non-decreasing.

\end{IEEEproof}

\end{document}